\documentclass[english]{sbrt}
\usepackage[english]{babel}
\usepackage[utf8]{inputenc}

\usepackage{tabularx}
\usepackage{booktabs}

\usepackage{graphicx}
\graphicspath{{.}{.\figs}}

\usepackage{graphicx} 
\usepackage{amsmath, amssymb,mathtools, stmaryrd, gensymb,algorithm,algpseudocode,float,subfloat,subfigure,csquotes}

\usepackage[%
style=ieee,
backend=biber, 
]{biblatex}

\defbibheading{bibliography}[\refname]{\section*{#1}}

\makeatletter
\g@addto@macro{\UrlBreaks}{\UrlOrds}
\makeatother

\usepackage{acro}[=v2]

\NewDocumentCommand{\acro}{m o m o}
{%
	\IfValueTF{#2}{%
		\IfValueTF{#4}{%
			\DeclareAcronym{#1}{short={#2},long={#3},#4}
		}{%
			\DeclareAcronym{#1}{short={#2},long={#3}}
		}
	}{%
		\IfValueTF{#4}{%
			\DeclareAcronym{#1}{short={#1},long={#3},#4}
		}{%
			\DeclareAcronym{#1}{short={#1},long={#3}}
		}
	}
}

\acro{2D}{two-dimensional}
\acro{3D}{three-dimensional}
\acro{4D}{four-dimensional}
\acro{6D}{six-dimensional}
\acro{1G}{first generation}
\acro{2G}{second generation}
\acro{3G}{third generation}
\acro{4G}{fourth generation}
\acro{5G}{fifth generation}
\acro{5GC}{5G core network}
\acro{5G-StoRM}{5G stochastic radio channel for dual mobility}
\acro{6G}{sixth generation}
\acro{3GPP}{3rd generation partnership project}
\acro{3GPP2}{3rd generation partnership project 2}
\acro{A2A}{air-to-air}
\acro{A2G}{air-to-ground}
\acro{AA}{Antenna Array}
\acro{AC}{Admission Control}
\acro{AD}{Attack-Decay}
\acro{ADC}{analog-to-digital converter}
\acro{AE}{Antenna Element}
\acro{AF}{Amplify and Forward}
\acro{ABS}{Almost Blank Subframe}
\acro{ACF}{Autocorrelation Function}
\acro{ADSL}{Asymmetric Digital Subscriber Line}
\acro{AHW}{Alternate Hop-and-Wait}
\acro{AI}{Artificial Intelligence}
\acro{AMC}{Adaptive Modulation and Coding}
\acro{AP}{Access Point}
\acro{APA}{Adaptive Power Allocation}
\acro{API}{Application Protocol Interface}
\acro{AT}{averaged time}
\acro{AR}{Autoregressive}
\acro{ARIMA}{Autoregressive Integrated Moving Average}
\acro{ARMA}{Autoregressive Moving Average}
\acro{ASE}{Average Squared Error}
\acro{ASC}{Adaptive Satisfaction Control}
\acro{ATES}{Adaptive Throughput-based Efficiency-Satisfaction Trade-Off}
\acro{AWGN}{Additive White Gaussian Noise}
\acro{BB}{Branch and Bound}
\acro{BC}{Branch and Cut}
\acro{BD}{block diagonalization}
\acro{BER}{Bit Error Rate}
\acro{BF}{Best Fit}
\acro{BL}{Bit Loading}
\acro{BLER}{BLock Error Rate}
\acro{BLPC-1}{Bit Loading with Power Constraint in Hop 1}
\acro{BLPC-2}{Bit Loading with Power Constraint in Hop 2}
\acro{BPC}{Binary Power Control}
\acro{BPSK}{Binary Phase-Shift Keying}
\acro{BRA}{Balanced Random Allocation}
\acro{BS}{base station}
\acro{BSP}{\acs*{BS} Placement}
\acro{CAP}{Combinatorial Allocation Problem}
\acro{CAPEX}{Capital Expenditure}
\acro{CB}{Contextual Bandit}
\acro{CBF}{Coordinated Beamforming}
\acro{CBR}{Constant Bit Rate}
\acro{CBS}{Class Based Scheduling}
\acro{CC}{Congestion Control}
\acro{CCL}{Common Cell List}
\acro{CDF}{Cumulative Distribution Function}
\acro{CDL}{Clustered Delay Line}
\acro{CDMA}{Code-Division Multiple Access}
\acro{CIR}{channel impulse response}
\acro{CH}{Channel Hardening}
\acro{CHO}{Conditional Handover}
\acro{C-RAN}{cloud-based Radio Access Network}
\acro{CL}{Closed Loop}
\acro{CLT}{central limit theorem}
\acro{CLPC}{Closed Loop Power Control} 
\acro{CN}{Core Network}       
\acro{CNR}{Channel-to-Noise Ratio}
\acro{CPA}{Cellular Protection Algorithm}
\acro{CPICH}{Common Pilot Channel}
\acro{CoMP}{Coordinated Multi-Point}
\acro{CQI}{Channel Quality Indicator}
\acro{CRE}{Cell Range Expansion}
\acro{CRM}{Constrained Rate Maximization}
\acro{CRN}{Cognitive Radio Network}
\acro{C-RNTI}{Cell Radio Network Temporary Identifier}
\acro{CRRM}{Centralized/Common Radio Resource Management}
\acro{CRS}{Cell-specific Reference Signal}
\acro{CS}{Coordinated Scheduling}
\acro{CSI}{Channel State Information}
\acro{CTS}{Clear to Send}
\acro{CU}{Central Unit}
\acro{CUE}{Cellular User Equipment}
\acro{CWND}{Congestion window size}
\acro{D2D}{Device-to-Device}
\acro{D2N}{Drone-to-Network}
\acro{DAA}{Drone Antenna Array}
\acro{DAG}{Directed Acyclic Graph}
\acro{DBS}{Drone Mounted Base Station}
\acro{DC}{Dual Connectivity}
\acro{DCA}{Dynamic Channel Allocation}
\acro{DE}{Differential Evolution}
\acro{DF}{Decode and Forward}
\acro{DFT}{Discrete Fourier Transform}
\acro{DIST}{Distance}
\acro{DL}{Downlink}
\acro{DMA}{Double Moving Average}
\acro{DMRS}{Demodulation Reference Signal}
\acro{D2DM}{\acs*{D2D} Mode}
\acro{DMS}{\acs*{D2D} Mode Selection}
\acro{DPC}{Dirty Paper Coding}
\acro{DRA}{Dynamic Resource Assignment}
\acro{DSA}{Dynamic Spectrum Access}
\acro{DSM}{Delay-based Satisfaction Maximization}
\acro{DU}{Distributed Unit}
\acro{E2E}{End-to-End}
\acro{ECC}{Electronic Communications Committee}
\acro{EDF}{Earliest Deadline First}
\acro{EE}{Energy Efficiency}
\acro{EFLC}{Error Feedback Based Load Control}
\acro{EI}{Efficiency Indicator}
\acro{e-ICIC}{Enhanced Inter-Cell Interference Coordination}
\acro{eMBB}{Enhanced Mobile Broadband}
\acro{eNB}{Evolved Node B}
\acro{EXP}{Exponential}
\acro{EPA}{Equal Power Allocation}
\acro{EPC}{Evolved Packet Core}
\acro{EPS}{Evolved Packet System}
\acro{E-UTRAN}{Evolved Universal Terrestrial Radio Access Network}
\acro{ES}{Exhaustive Search}
\acro{FCP}{Fundamental Counting Principle}
\acro{FCA}{Flow Control Algorithm}
\acro{FD}{Full-Duplex}
\acro{FDD}{Frequency Division Duplex}
\acro{FDM}{Frequency Division Multiplexing}
\acro{FDMA}{Frequency Division Multiple Access}
\acro{FER}{Frame Erasure Rate}
\acro{FIB}{Full \acs*{IRS} \acs*{BD}}
\acro{FIFO}{First In First Out}
\acro{FF}{Fast Fading}
\acro{FRS}[FS]{Fast-RAT Scheduling}
\acro{FS}{Fast Switching}
\acro{FSB}{Fixed Switched Beamforming}
\acro{FST}{Fixed \acs*{SNR} Target}
\acro{FT}{Fourier transform}
\acro{FTP}{File Transfer Protocol}
\acro{GA}{Genetic Algorithm}
\acro{GAP}{Generalized Assignment Problem}
\acro{GAP-MQ}{Generalized Assignment Problem with Minimum Quantities}
\acro{GATES}{Generalized Adaptive Throughput-based Efficiency-Satisfaction Trade-Off}
\acro{GBR}{Guaranteed Bit Rate}
\acro{GLR}{Gain to Leakage Ratio}
\acro{gNB}{gNode B}
\acro{GOS}{Generated Orthogonal Sequence}
\acro{GP}{Gaussian Process}
\acro{GPL}{GNU General Public License}
\acro{GPS}{Global Positioning System}
\acro{GRP}{Grouping}
\acro{GSM}{Global System for Mobile Communications}
\acro{GTEL}{Wireless Telecommunications Research Group}
\acro{HARQ}{Hybrid Automatic Repeat Request}
\acro{HBF}{Hybrid Beamforming}
\acro{HCPP}{Hardcore Point Process}
\acro{HD}{High Definition}
\acro{HetNet}{Heterogeneous Network}
\acro{HH}{Hughes-Hartogs}
\acro{HardH}[HH]{Hard Handover}
\acro{HMS}{Harmonic Mode Selection}
\acro{HO}{Handover}
\acro{HOL}{Head Of Line}
\acro{HPBW}{Half Power Beamwidth}
\acro{HSDPA}{High Speed Downlink Packet Access}
\acro{HSR}{high-Speed Railway}
\acro{HSPA}{High Speed Packet Access}
\acro{HTTP}{HyperText Transfer Protocol}
\acro{IAB}{Integrated Access and Backhaul}
\acro{ICMP}{Internet Control Message Protocol} 
\acro{ICI}{Intercell Interference}
\acro{ICIC}{Inter-Cell Interference Coordination}
\acro{ID}{Identification}
\acro{i.i.d.}{independent and identically distributed}
\acro{IETF}{Internet Engineering Task Force}
\acro{IPC}{Individual Power Constraint}
\acro{UID}{Unique Identification}
\acro{IID}{Independent and Identically Distributed}
\acro{IIR}{Infinite Impulse Response}
\acro{ILP}{Integer Linear Problem}
\acro{IMT}{International Mobile Telecommunications}
\acro{INV}{Inverted Norm-based Grouping} 
\acro{IoT}{Internet of Things}
\acro{IP}{Internet Protocol}
\acro{IPv6}{Internet Protocol Version 6}
\acro{IRA}{Integrated Resource Allocation}
\acro{IRS}{intelligent reflecting surface}
\acro{ISD}{Inter-Site Distance}
\acro{ISI}{Inter Symbol Interference}
\acro{ISM}{Industrial, Scientific and Medical}
\acro{ITU}{International Telecommunication Union}
\acro{JOAS}{Joint Opportunistic Assignment and Scheduling}
\acro{JOS}{Joint Opportunistic Scheduling}
\acro{JP}{Joint Processing}
\acro{JRAPAP}{Joint RB Assignment and Power Allocation Problem}
\acro{JS}{Jump-Stay}
\acro{JSM}{Joint Satisfaction Maximization}
\acro{KKT}{Karush-Kuhn-Tucker}
\acro{KPI}{key performance indicator}
\acro{LAC}{Link Admission Control}
\acro{LA}{Link Adaptation}
\acro{LBS}{Location Based Service}
\acro{LC}{Load Control}
\acro{LOS}{line of sight}
\acro{LP}{Linear Programming}
\acro{LSTM}{Long Short-Term Memory}
\acro{LTE}{Long Term Evolution}
\acro{LTE-A}{\acs*{LTE}-Advanced}
\acro{LTE-Advanced}{\ac{LTE-A}}
\acro{LSP}{Large Scale Parameter}
\acro{MA}{Moving Average}
\acro{MADRDPG}{Multi-Agent Deep Recurrent Deterministic Policy Gradient}
\acro{MeNB}{Master \acs*{eNB}}
\acro{M2M}{Machine-to-Machine}
\acro{MAC}{Medium Access Control}
\acro{MANET}{Mobile Ad hoc Network}
\acro{MEDS}{Method of Exact Doppler Spread}
\acro{MC}{Modular Clock}
\acro{MCP}{Minimal Cost Power}
\acro{MCS}{Modulation and Coding Scheme}
\acro{MDB}{Measured Delay Based}
\acro{MDI}{Minimum \acs*{D2D} Interference}
\acro{MDSM}{Modified Delay-based Satisfaction Maximization}
\acro{MDU}{Max-Delay-Utility}
\acro{METIS}{Mobile and Wireless Communications Enablers for the Twenty-twenty Information Society \acs*{5G}}
\acro{MF}{Matched Filter}
\acro{MFAP}{Mobile Femtocell Access Point}
\acro{MG}{Maximum Gain}
\acro{MH}{Multi-Hop}
\acro{MILP}{Mixed Integer Linear Programming}
\acro{MIMO}{multiple input multiple output}
\acro{MINLP}{Mixed Integer Nonlinear Programming}
\acro{MIP}{Mixed Integer Programming}
\acro{MISO}{Multiple Input Single Output}
\acro{MIT}{Mobility Interruption Time}
\acro{ML}{Machine Learning}
\acro{MLWDF}{Modified Largest Weighted Delay First}
\acro{MME}{Mobility Management Entity}
\acro{MMF}{Max-Min Fairness}
\acro{mmMAGIC}{Millimetre-Wave Based Mobile Radio Access Network for Fifth Generation Integrated Communications}
\acro{MMRP}{Maximizing the Minimal Residual Power}
\acro{MMRP-LB}{Maximizing the Minimal Residual Power with Lower Bound}
\acro{MMSE}{Minimum Mean Square Error}
\acro{mMTC}{Massive Machine-Type Communications}
\acro{mmWave}{millimeter wave}
\acro{MN}{Master Node}
\acro{MOS}{Mean Opinion Score}
\acro{MPF}{Multicarrier Proportional Fair}
\acro{MPRP}{Maximization of the Product of the Residual Powers}
\acro{MRA}{Maximum Rate Allocation}
\acro{MR}{Maximum Rate}
\acro{MRC}{Maximum Ratio Combining}
\acro{MRN}{Mobile Relay Node}
\acro{MRT}{Maximum Ratio Transmission}
\acro{MRUS}{Maximum Rate with User Satisfaction}
\acro{MS}{Mode Selection}
\acro{MSE}{Mean Squared Error}
\acro{MCG}{Master Cell Group} 
\acro{MSI}{Multi-Stream Interference}
\acro{MTC}{Machine-Type Communication}
\acro{IMS}{\acs*{IP} Multimedia Subsystem}
\acro{MT}{Mobile Termination}
\acro{MTSI}{Multimedia Telephony Services over \acs*{IMS}}
\acro{MTSM}{Modified Throughput-based Satisfaction Maximization}
\acro{MU-MIMO}{multi-user \acs*{MIMO}}
\acro{mMIMO}{massive \acs*{MIMO}}
\acro{MU}{Multi-User}
\acro{Multi-CUT}{Multi-Cell and Multi-User and Multi-Tier}
\acro{NAS}{Non-Access Stratum}
\acro{NB}{Node B}
\acro{nBS}{Neighbor Base Station}
\acro{NCL}{Neighbor Cell List}
\acro{NGC}{Next Generation Core Network}
\acro{NLP}{Nonlinear Programming}
\acro{NLOS}{Non-Line of Sight}
\acro{NMSE}{Normalized Mean Square Error}
\acro{NN}{Neural Network}
\acro{NOMA}{Non-Orthogonal Multiple Access}
\acro{NORM}{Normalized Projection-based Grouping}
\acro{NP}{Non-Polynomial Time}
\acro{NR}{New Radio}
\acro{NRT}{Non-Real Time}
\acro{NSA}{Non-Stand-Alone}
\acro{NSPS}{National Security and Public Safety Services}
\acro{OBF}{Opportunistic Beamforming}
\acro{OFDMA}{Orthogonal Frequency Division Multiple Access}
\acro{OFDM}{Orthogonal Frequency Division Multiplexing}
\acro{OFPC}{Open Loop with Fractional Path Loss Compensation}
\acro{O2I}{Outdoor-to-Indoor}
\acro{OL}{Open Loop}
\acro{OLPC}{Open-Loop Power Control}
\acro{OL-PC}{Open-Loop Power Control}
\acro{OM}[O\&M]{Operational \& Maintenance}
\acro{OMA}{Orthogonal Multiple Access}
\acro{OPEX}{Operational Expenditure}
\acro{ORA}{Orthogonal Resource Allocation}
\acro{ORB}{Orthogonal Random Beamforming}
\acro{JO-PF}{Joint Opportunistic Proportional Fair}
\acro{OSI}{Open Systems Interconnection}
\acro{PA}{Power Allocation}
\acro{PACF}{Partial Autocorrelation Function}
\acro{PAIR}{\acs*{D2D} Pair Gain-based Grouping}
\acro{PAPR}{Peak-to-Average Power Ratio}
\acro{P2P}{Peer-to-Peer}
\acro{PBCH}{Physical Broadcast Channel}
\acro{PBS}{Pico Base Station}        
\acro{PC}{Power Control}
\acro{PCI}{Physical Cell \acs*{ID}}
\acro{PDCP}{Packet Data Convergence Protocol}
\acro{PDF}{Probability Density Function}
\acro{PDU}{Protocol Data Unit}
\acro{PER}{Packet Error Rate}
\acro{PF}{Proportional Fair}
\acro{PGRA}{Probabilistic Graph based Resource Allocation}
\acro{P-GW}{Packet Data Network Gateway}
\acro{PHY}{Physical}
\acro{PIB}{Partial \acs*{IRS} \acs*{BD}}
\acro{PL}{Pathloss}
\acro{PLR}{Packet Loss Ratio}
\acro{PRABE}{Power and Resource Allocation Based on Quality of Experience}
\acro{PRB}{Physical Resource Block}
\acro{PROJ}{Projection-based Grouping}
\acro{PSD}{Power Spectral Density}
\acro{PSDe}{positive semi-definite}
\acro{ProSe}{Proximity Services}
\acro{PS}{Packet Scheduling}
\acro{PSO}{Particle Swarm Optimization}
\acro{PSS}{Primary Synchronization Signal}
\acro{PTAS}{Polynomial-Time Approximation Scheme}
\acro{PZF}{Projected Zero-Forcing}
\acro{QAM}{Quadrature Amplitude Modulation}
\acro{QHMLWDF}{Queue-HOL-MLWDF}
\acro{QoE}{Quality of Experience}
\acro{QoS}{Quality of Service}
\acro{QPSK}{Quadri-Phase Shift Keying}
\acro{QSM}{Queue-based Satisfaction Maximization}
\acro{QuaDRiGa}{QUAsi Deterministic RadIo channel GenerAtor}
\acro{RACH}{Random Access}
\acro{RAISES}{Reallocation-based Assignment for Improved Spectral Efficiency and Satisfaction}
\acro{RAN}{Radio Access Network}
\acro{RA}{Resource Allocation}
\acro{RAP}{RB Assignment Problem}
\acro{RAT}{Radio Access Technology}[long-plural-form={Radio Access Technologies}]
\acro{RATE}{Rate-based}
\acro{RAU}{Remote Antenna Unit}
\acro{RB}{Resource Block}
\acro{RBG}{Resource Block Group}
\acro{REF}{Reference Grouping}
\acro{RET}{Remote Electrical Tilt}
\acro{RF}{radio frequency}
\acro{RL}{Reinforcement Learning}
\acro{RLC}{Radio Link Control}
\acro{RLF}{Radio Link Failure}
\acro{RM}{Rate Maximization}
\acro{RMa}{Rural Macro}
\acro{RMJ-SNR}{Region of the Minimum Joint SNR}
\acro{RMEC}{Rate Maximization under Experience Constraints}
\acro{RMSE}{root-mean-square error}
\acro{RNC}{Radio Network Controller}
\acro{RND}{Random Grouping}
\acro{RNN}{Recurrent Neural Network}
\acro{RRA}{Radio Resource Allocation}
\acro{RRM}{Radio Resource Management}
\acro{RSCP}{Received Signal Code Power}
\acro{RSRP}{Reference Signal Received Power}
\acro{RSRQ}{Reference Signal Received Quality}
\acro{RR}{Round Robin}
\acro{RRC}{Radio Resource Control}
\acro{RSSI}{Received Signal Strength Indicator}
\acro{RT}{Real Time}
\acro{RTS}{Request to Send}
\acro{RU}{Resource Unit}
\acro{RUNE}{RUdimentary Network Emulator}
\acro{RV}{Random Variable}
\acro{Rx}{Receiver}
\acro{RZF}{Regularized Zero-Forcing}
\acro{SA}{Subcarrier Assignment}
\acro{SAC}{Session Admission Control}
\acro{sBS}{Serving Base Station}
\acro{SC}{Small Cell}
\acro{SCon}[SC]{Single Connectivity}
\acro{SCG}{Secondary Cell Group}
\acro{SCM}{Spatial Channel Model}
\acro{SCS}{Subcarrier Spacing}
\acro{SC-FDMA}{Single Carrier - Frequency Division Multiple Access}
\acro{SCRV}{Spatially Consistent Random Variable}
\acro{SD}{Soft Dropping}
\acro{S-D}{Source-Destination}
\acro{SDPC}{Soft Dropping Power Control}
\acro{SDMA}{Space-Division Multiple Access}
\acro{SDR}{Software Defined Radio}
\acro{SE}{spectral efficiency}
\acro{SF}{Shadow Fading}
\acro{SMDP}{Semi-Markov	Decision Problem}
\acro{SeNB}{Secondary \acs*{eNB}}
\acro{SER}{symbol error rate}
\acro{SES}{Simple Exponential Smoothing}
\acro{S-GW}{Serving Gateway}
\acro{SINR}{signal-to-interference-plus-noise-ratio}
\acro{SLNR}{signal to Leakage-plus-Noise Ratio}
\acro{SNN}{strictly non-negative}
\acro{SI}{Satisfaction Indicator}
\acro{SIP}{Session Initiation Protocol}
\acro{SISO}{Single Input Single Output}
\acro{SIMO}{Single Input Multiple Output}
\acro{SIR}{Signal to Interference Ratio}
\acro{SM}{Subcarrier Matching}
\acro{SMA}{Simple Moving Average}
\acro{SMSE}{spatial-mean-square error}
\acro{SN}{Secondary Node}
\acro{SNR}{signal-to-noise-ratio}
\acro{SON}{Self Organizing Networks}
\acro{SoA}{state-of-the-art}
\acro{SoS}{Sum-of-Sinusoids}
\acro{SORA}{Satisfaction Oriented Resource Allocation}
\acro{SORA-NRT}{Satisfaction-Oriented Resource Allocation for Non-Real Time Services}
\acro{SORA-RT}{Satisfaction-Oriented Resource Allocation for Real Time Services}
\acro{SP}{Signal Processing}
\acro{SPF}{Single-Carrier Proportional Fair}
\acro{SRA}{Sequential Removal Algorithm}
\acro{SRB1}{Signaling Radio Bearer~1}
\acro{SRE}{smart radio environment}
\acro{SRS}{Sounding Reference Signal}
\acro{SSB}{Synchronization Signal Block}
\acro{SSP}{Small Scale Parameter}
\acro{SSS}{Secondary Synchronization Signal}
\acro{ST}{Spanning Tree}
\acro{STTD}{Space Time Transmit Diversity}[long-plural-form={Space Time Transmit Diversities}]
\acro{SU-MIMO}{Single-User Multiple Input Multiple Output}
\acro{SU}{Single-User}
\acro{tBS}{Target Base Station}
\acro{SVD}{singular value decomposition}
\acro{TCP}{Transmission Control Protocol}
\acro{TDD}{Time Division Duplex}
\acro{TDMA}{Time Division Multiple Access}
\acro{TETRA}{Terrestrial Trunked Radio}
\acro{TP}{Transmit Power}
\acro{TPr}{Trend Prediction}
\acro{TPC}{Total Power Constraint}
\acro{TR}{Technical Report}
\acro{TTI}{Transmission Time Interval}
\acro{TTR}{Time-To-Rendezvous}
\acro{TTT}{Time-To-Trigger}
\acro{TS}{Time Series}
\acro{TSM}{Throughput-based Satisfaction Maximization}
\acro{TU}{Typical Urban}
\acro{TV}{Television}
\acro{TVWS}{\acs*{TV} White Space}
\acro{Tx}{Transmitter}
\acro{UAV}{Unmanned Aerial Vehicle}
\acro{UABSC}{User-Assisted Bearer Split Control}
\acro{UDP}{User Datagram Protocol}
\acro{USIM}{Universal Subscriber Identity Module}
\acro{UDN}{ultra-dense networks}
\acro{UE}{user equipment}
\acro{UBA}{\ac{UE}-\ac{BS} association}
\acro{UEPS}{Urgency and Efficiency-based Packet Scheduling}
\acro{UFC}{Federal University of Cear\'{a}}
\acro{ULA}{uniform linear array}
\acro{UL}{Uplink}
\acro{USRP}{Universal Software Radio Peripheral}
\acro{UMa}{urban macro}
\acro{UMi}{Urban Micro}
\acro{UMTS}{Universal Mobile Telecommunications System}
\acro{UPN}{User Provided Network}
\acro{URA}{uniform rectangular array}
\acro{URI}{Uniform Resource Identifier}
\acro{URLLC}{Ultra-Reliable Low-Latency Communications}
\acro{URM}{Unconstrained Rate Maximization}
\acro{VBR}{Variable Bit Rate}
\acro{VET}{Variable Electrical Tilt}
\acro{VR}{Virtual Resource}
\acro{VoIP}{Voice over \acs*{IP}}
\acro{WCDMA}{Wideband Code Division Multiple Access}
\acro{WF}{Water-filling}
\acro{WID}{wireless Infrastructure Drone}
\acro{Wi-Fi}{Wireless Fidelity}
\acro{WiMAX}{Worldwide Interoperability for Microwave Access}
\acro{WINNER}{Wireless World Initiative New Radio}
\acro{WLAN}{Wireless Local Area Network}
\acro{WMPF}{Weighted Multicarrier Proportional Fair}
\acro{WP}{Work Package}
\acro{WPF}{Weighted Proportional Fair}
\acro{WSN}{Wireless Sensor Network}
\acro{WWW}{World Wide Web}
\acro{WFQ}{Weighted Fair Queuing}
\acro{XIXO}{(Single or Multiple) Input (Single or Multiple) Output}
\acro{XPR}{Cross-Polarization Power Ratio}
\acro{ZF}{zero-forcing}
\acro{ZMCSCG}{Zero Mean Circularly Symmetric Complex Gaussian}

\usepackage{siunitx}
\newcommand{\SecRef}[2][]{Section#1~\ref{#2}}
\newcommand{\FigRef}[2][]{Fig.#1~\ref{#2}}

\newcommand{\TabRef}[2][]{Table#1~\ref{#2}}

\begin{document}

\title{Interference mitigation with block diagonalization for IRS-aided MU-MIMO communications}

\author{Wilker de O. Feitosa, Igor M. Guerreiro, Fco. Rodrigo P. Cavalcanti, Tarcisio F. Maciel,\\ Maria Clara R. Lobão, Fazal-E-Asim, Behrooz Makki and Gábor Fodor.	
\thanks{This work was supported in part by Ericsson Research, Technical Cooperation Contract UFC.51, in part by the Brazilian National Council for Scientific and Technological Development (CNPq), 
	in part by the Coordenação de Aperfeiçoamento de Pessoal de Nível Superior - Brasil (CAPES) - Finance Code 001, and in part by CNPq  306845/2020-2.
}
\thanks{Wilker de O. Feitosa, Igor M. Guerreiro, Fco. Rodrigo P. Cavalcanti, Tarcisio F. Maciel, Maria Clara R. Lobão  and Fazal-E-Asim are with the Wireless Telecommunications Research Group (GTEL), Federal University of Cear\'a (UFC), Fortaleza, Brazil. E-mails: \{wilker, igor, rodrigo, maciel, clara, fazalasim\}@gtel.ufc.br. Behrooz Makki and Gábor Fodor are with Ericsson Research. (e-mails: behrooz.makki@ericsson.com, gabor.fodor@ericsson.com).
}%
}

\maketitle

\markboth{XLI BRAZILIAN SYMPOSIUM ON TELECOMMUNICATIONS AND SIGNAL PROCESSING - SBrT 2023, OCTOBER 08--11, 2023, S\~{A}O JOSÉ DOS CAMPOS, SP}{}

\begin{abstract}
This work investigates interference mitigation techniques in multi-user multiple input multiple output (MU-MIMO) Intelligent Reflecting Surface (IRS)-aided networks, focusing on the base station end. Two methods of precoder design based on block diagonalization are proposed. The first method does not consider the interference caused by the IRS, seeking to mitigate only the multi-user interference. The second method mitigates both the IRS-caused interference and the multi-user interference. A comparison between both methods within an no-IRS MU-MIMO network with strong direct links is provided. The results show that, although in some circumstances IRS interference can be neglected, treating it can improve system capacity and provide higher spectral efficiency. 
\end{abstract}
\begin{keywords}
MU-MIMO, Interference Mitigation, Block Diagonalization, Intelligent Reflecting Surfaces.
\end{keywords}

\section{Introduction}

The \ac{6G} of cellular networks is expected to present significant advances in terms of system capacity, energy efficiency, number of supported users and~\ac{SE} compared to the~\ac{5G}~\cite{Zhang2019}. %
To accomplish this goal, new physical layer technologies are investigated to take more advantage of the propagation features of the environment, such as~\ac{IRS}~\cite{Hassouna2023}, sub-Terahertz bands, distributed~\ac{MIMO}, among others~\cite{rajatheva2021}. %

\Ac{MU-MIMO}, as a well-established key technology in mobile wireless systems due to its advantages in spatial diversity and multiplexing, will also play an important role in \ac{6G}, where its beamforming gains and improvements in~\ac{SE} are desired and enhanced when combined with the aforementioned technologies. %
One of the challenges in \ac{MU-MIMO} systems is to deal with multi-user interference. %
Various methods for mitigating interference on the receiver end, as well as at the transmitter, have been developed over the past years, for instance, the design of robust decoding and precoding filters using methods like \ac{ZF}~\cite{Joham2005} and \ac{BD}~\cite{Spencer2004}. %

In~\ac{5G} and beyond systems, the use of \ac{mmWave} bands is highly desirable due to the large bandwidth available in the spectrum. %
However, since high frequencies are bound to severe pathloss and penetration loss, communication in those bands is much more susceptible to blockage and poor link conditions without a~\ac{LOS} component. %
Such effects can even interrupt the connection, thus limiting the capacity of the system. %
In order to overcome these limitations, the use of~\ac{mmWave} is usually combined with other technologies like~\ac{mMIMO} and ultra-dense networks~\cite{Busari2018}. %
In these circumstances, the concept of~\ac{SRE} can be introduced, which states that the wireless environment can be partially turned into an optimization variable that, jointly with transmitter and receiver properties, can be used to maximize the overall network performance~\cite{Renzo2020}. %

The concept of~\acp{IRS} is a candidate enabler for~\ac{SRE} since it can well modify the environment. %
For instance,~\ac{IRS} can create a virtual~\ac{LOS} component, higher-rank channels, or attenuate undesired signals~\cite{Wu2021}. %
\acp{IRS} also actuate as antenna arrays, improving signal quality by applying beamforming to the desired signal. %
Thus,~\ac{SRE} aided by~\acp{IRS} can be leveraged to diminish the effects of propagation losses, improve the coverage and increase the~\ac{SE} by optimizing the environment between transmitter and receiver to achieve better link conditions~\cite{Gong2020}. %

Considering the high pathloss and blockage probability at~\ac{mmWave} bands, the use of~\acp{IRS} can provide beamforming gains and additional paths to users under poor propagation conditions, thus improving the capacity of~\ac{MU-MIMO} systems. %
Nonetheless, it is important to notice that even a fully passive~\ac{IRS}, i.e., an IRS without~\ac{RF} chains, can introduce interference to untargeted users and~\acp{BS}, e.g., other nodes nearby the intended user. %
This raises the question about the need to mitigate such interference and the means to do it. %

To manage the interference introduced by \acp{IRS}, the authors in~\cite{Ning2022} employ an orthogonalization scheme based on~\ac{BD} in a~\ac{MU-MIMO} scenario. %
However, their \ac{BD} approach demands the use of least one~\ac{IRS} per user.%
The authors in~\cite{Zheng2022} also address this problem by minimizing the~\ac{SER} using 1-bit~\acp{ADC} on the~\ac{BS} side. %

In this paper, we study the problem of interference management in~\ac{IRS}-assisted \ac{MU-MIMO} networks with a single~\ac{IRS}. %
We propose two precoding methods based on~\ac{BD}. %
Compared to the solution in~\cite{Ning2022}, our proposed methods mitigate interference caused by \acp{IRS} not only by being less greedy in terms of computational complexity, but also using fewer~\ac{RF} chains at both the transmitter and the receiver. %
The first method considers the interference caused by the~\ac{IRS} as negligible and focuses only on multi-user interference mitigation. %
The second method takes both types of interference into account and also uses part of the~\ac{IRS} signal towards untargeted users as useful signal. %
The results show that interference can be mitigated with both methods, and in particular, the second method presents the highest~\ac{SE} in comparison with the other method and the state-of-the-art.

\section{System model}\label{SEC:SystemModel}

\begin{figure}[!t]
	\centering
	\includegraphics[width=0.7\columnwidth]{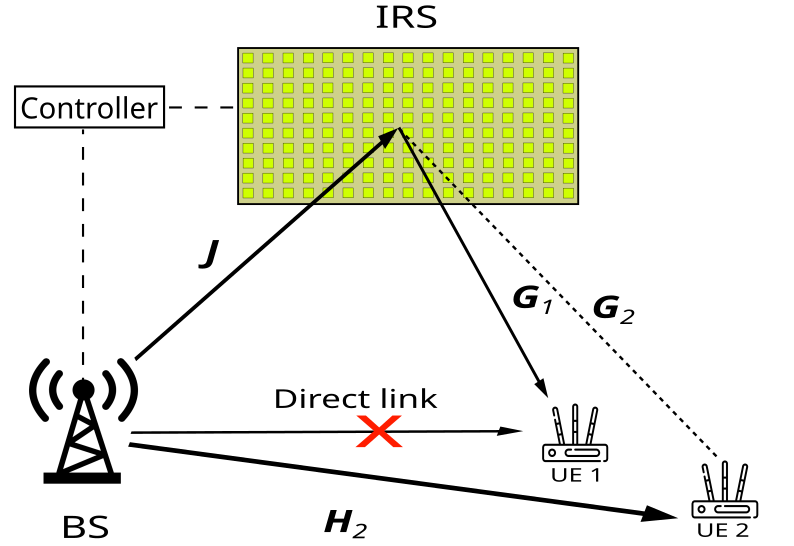}
	\caption{Multi-user MIMO-IRS assisted systems}
	\label{FIG:single-cell}
\end{figure}

Consider the downlink of a system composed of a \ac{BS} with $M$ antennas serving simultaneously two co-channel \acp{UE}, as shown in~\FigRef{FIG:single-cell}. %
\ac{UE}1 is equipped with $Q$ antennas and has no direct path to the \ac{BS} due to, e.g., a strong blockage effect. Therefore, the \ac{BS} serves \ac{UE}1 \emph{via} an \ac{IRS} with $N$ reflecting elements. %
\ac{UE}2, with $P$ antennas, has a strong direct link to the \ac{BS}; thus, the use of the IRS for UE2 is optional. %
The~\ac{IRS} phase-shift controller is assumed to work ideally with the~\ac{BS}. %
While serving \ac{UE}1 \textit{via} IRS,~\ac{UE}2 receives from the \ac{BS} an unintended signal due to beam leakage, whose intensity depends on the propagation conditions and on the \ac{BS} transmit power. %

The received signal model is given as:
\begin{equation}\label{EQ:MIMOMatricial_step}
	\begin{bmatrix}
		\mathbf{y}_1 \\ \mathbf{y}_2
	\end{bmatrix} = 
	\begin{bmatrix}
		\mathbf{W}_1^\text{H} & \mathbf{0}  \\
		\mathbf{0} & \mathbf{W}_2^\text{H}
	\end{bmatrix}
	\begin{bmatrix}
		\bar{\mathbf{H}}_1   \\
		\bar{\mathbf{H}}_2
	\end{bmatrix}
	\begin{bmatrix}
		\mathbf{F}_1 &
		\mathbf{F}_2
	\end{bmatrix}
	\begin{bmatrix}
		\mathbf{x}_1 \\
		\mathbf{x}_2
	\end{bmatrix} + \begin{bmatrix}
		\tilde{\mathbf{n}}_1 \\
		\tilde{\mathbf{n}}_2
	\end{bmatrix}, 
\end{equation}
where $\mathbf{W}_1 \in \mathbb{C}^{Q \times N_s^{(1)}}$ represents the combiner for \ac{UE}1, $\mathbf{F}_1 \in \mathbb{C}^{M \times N_s^{(1)}}$ is the digital baseband precoder for \ac{UE}1, $\mathbf{n}_1 \sim \mathcal{CN}\left(\mathbf{0},\sigma_1^2\mathbf{I}_{Q}\right)$ is the circularly symmetric additive white Gaussian noise vector with variance $\sigma_1^2$ and $\mathbf{x}_1 \in \mathbb{C}^{N_s^{(1)} \times 1}$ is the data vector for \ac{UE}1. %
Similarly, $\mathbf{W}_2 \in \mathbb{C}^{P \times N_s^{(2)}}$ is the combiner for \ac{UE}2, $\mathbf{F}_2 \in \mathbb{C}^{M \times N_s^{(2)}}$ is the digital baseband precoder for \ac{UE}2 and $\mathbf{n}_2 \sim \mathcal{CN}\left(\mathbf{0},\sigma_2^2\mathbf{I}_{P}\right)$ is the circularly symmetric additive white Gaussian noise vector with variance $\sigma_2^2$, and finally, $\mathbf{x}_2 \in \mathbb{C}^{N_s^{(2)} \times 1}$ is the data vector for \ac{UE}2. %
At last, $\mathbf{I}_{a}$ denotes the $a \times a$ identity matrix and $\mathbf{0}$ is a vector of zeros of proper size. %
\subsection{Signal Model}

Given propagation scenario depicted in Fig.~\ref{FIG:single-cell}, the channel for \ac{UE}1 can be defined as $\bar{\mathbf{H}}_{1} = \mathbf{G}_{1}\boldsymbol{\Omega}\mathbf{J} \in \mathbb{C}^{Q \times M}$, hence, the complete equation of the received signal for \ac{UE}1 is written as %
\begin{equation}\label{EQ:y1CompleteEquation}
  \begin{split}
    \mathbf{y}_1 =& \mathbf{W}_1^\text{H}\mathbf{G}_1\boldsymbol{\Omega}\mathbf{J} \mathbf{F}_1\mathbf{x}_1 + \\ %
                  &\underbrace{\mathbf{W}_1^\text{H}\mathbf{G}_1\boldsymbol{\Omega}\mathbf{J} \mathbf{F}_2\mathbf{x}_2}_{\textbf{interference}} +  \underbrace{\mathbf{W}_1^\text{H}\mathbf{n}_1}_{\tilde{\mathbf{n}}_1} \in \mathbb{C}^{N_s^{(1)} \times 1},
  \end{split} %
\end{equation}
where $\mathbf{J} \in \mathbb{C}^{N \times M}$ is the channel between the BS and the IRS, %
$\mathbf{G}_1\in \mathbb{C}^{Q \times N}$ is the channel between the IRS and the \ac{UE}1 and $\boldsymbol{\Omega}=\text{diag}\left\{\boldsymbol{\omega}\right\}$, %
$\boldsymbol{\omega} \in \mathbb{C}^{N \times 1}$, is the IRS phase-shift vector.
Based on \eqref{EQ:y1CompleteEquation}, the \ac{SINR} $\gamma_1$ for \ac{UE}1 is given by 
\begin{equation}\label{EQ:R1} 
	\gamma_1 = \mathrm{tr}\left[\mathbf{W}_1^\text{H}\mathbf{G}_1\boldsymbol{\Omega}\mathbf{J}\mathbf{F}_1\mathbf{F}_1^\text{H}\mathbf{J}^\text{H}\boldsymbol{\Omega}^\text{H}\mathbf{G}_1^\text{H}\mathbf{W}_1\mathbf{R}_1^{-1}\right]\,,
\end{equation}
where $\mathbf{R}_1=  \sigma_1^2 \mathbf{I}_{N_s^{(1)}} + \mathbf{W}_1^\text{H}\bar{\mathbf{H}}_{1}\mathbf{F}_2\mathbf{F}_2^\text{H}\bar{\mathbf{H}}_{1}^\text{H}\mathbf{W}_1$, and $\mathrm{tr}[\cdot]$ denotes the trace operator.

Likewise, the channel for \ac{UE}2 can be defined as $\bar{\mathbf{H}}_2 = \mathbf{H}_2 + \mathbf{G}_2\boldsymbol{\Omega}\mathbf{J} \in \mathbb{C}^{P \times M}$. Therefore, the received signal for \ac{UE}2 from~\eqref{EQ:MIMOMatricial_step} is given by 
\begin{equation}\label{EQ:y2CompleteEquation}
  \begin{split}
    &\mathbf{y}_2 = \mathbf{W}_2^\text{H}\mathbf{H}_2\mathbf{F}_2\mathbf{x}_2 + \mathbf{W}_2^\text{H}\mathbf{G}_2\boldsymbol{\Omega}\mathbf{J}\mathbf{F}_2\mathbf{x}_2 + \\
    &\underbrace{\mathbf{W}_2^\text{H}\mathbf{H}_2\mathbf{F}_1\mathbf{x}_1 + \mathbf{W}_2^\text{H}\mathbf{G}_2\boldsymbol{\Omega}\mathbf{J}\mathbf{F}_1\mathbf{x}_1}_{\textbf{interference}} + \underbrace{\mathbf{W}_2^\text{H}\mathbf{n}_2}_{\tilde{\mathbf{n}}_2} \in \mathbb{C}^{N_s^{(2)} \times 1},
  \end{split}
\end{equation}
where $\mathbf{H}_2 \in \mathbb{C}^{P \times M}$ is the direct channel between the BS and the \ac{UE}2, %
$\mathbf{G}_2 \in \mathbb{C}^{P \times N}$ is the leakage channel between the IRS and \ac{UE}2. %
Now let $\mathbf{R}_2 = \sigma_2^2 \mathbf{I}_{N_s^{(2)}} + \mathbf{W}_2^\text{H}\bar{\mathbf{H}}_{2}\mathbf{F}_1\mathbf{F}_1^\text{H}\bar{\mathbf{H}}_{2}^\text{H}\mathbf{W}_2.$ 
Then, based on~\eqref{EQ:y2CompleteEquation}, the \ac{SINR} for \ac{UE}2 is calculated by~\eqref{EQ:R2}, given on top of the next page. %

The key performance indicators used for comparing the techniques are the~\ac{SE} and the sum~\ac{SE}. The~\ac{SE} can be calculated as $\epsilon_j = \log_{2}\det[\mathbf{I}_{N_s} + \gamma_{j}], j\in\{1,2\}$, and the sum~\ac{SE} is defined as $ \epsilon_\text{sum} = \sum_{j\in\{1,2\}} \epsilon_j$.

\begin{figure*}[!t]
  \normalsize
  \begin{equation}\label{EQ:R2}
	  \gamma_2 = \mathrm{tr}\left[ (\mathbf{W}_2^\text{H}\mathbf{H}_2\mathbf{F}_2\mathbf{F}_2^\text{H}\mathbf{H}_2^\text{H}\mathbf{W}_2 + \mathbf{W}_2^\text{H}\mathbf{G}_2\boldsymbol{\Omega}\mathbf{J}\mathbf{F}_2\mathbf{F}_2^\text{H}\mathbf{J}^\text{H}\boldsymbol{\Omega}^\text{H}  \mathbf{G}_2^\text{H}\mathbf{W}_2)\mathbf{R}_2^{-1}\right] 
   \end{equation}
  
  \hrulefill
   \vspace*{-5pt}
\end{figure*}

It is also important to notice that~\ac{mmWave} and Terahertz systems tend to have fewer \ac{RF} chains in their configurations, due to their massive number of antennas. %
In this paper, all hybrid beamforming is done considering that the number of \ac{RF} chains is smaller than the number of antennas~\cite{Zhang2020}. %

\subsection{Propagation Model}

Considering the scenario illustrated in \FigRef{FIG:single-cell}, our adopted channel model~\cite{HeathLozano2018} is given by
\begin{equation}\label{EQ:GeneralChannelModel}
  \begin{split}
    \mathbf{H}_{r,t} &= \sqrt{\cfrac{K}{K+1}}\;A_{0}\mathbf{a}_r\left(\theta_{r,0},\phi_{r,0}\right)\mathbf{a}^\text{T}_t\left(\theta_{t,0},\phi_{t,0}\right)  + \\
                     &\sqrt{\cfrac{1}{K+1}}\left(\cfrac{1}{\sqrt{\smash[b]{\vphantom{|}}\smash[b]{\vphantom{|}}S}}\sum_{s=1}^{S}A_s\mathbf{a}_r\left(\theta_{r,s},\phi_{r,s}\right)\mathbf{a}^\text{T}_t\left(\theta_{t,s},\phi_{t,s}\right)\right),
  \end{split}
\end{equation}
in which $K$ is the Rician K-factor; $\mathbf{a}_r\left(\theta_{r,s},\phi_{r,s}\right)$ and $\mathbf{a}_t\left(\theta_{t,s},\phi_{t,s}\right)$ represent the steering vectors at the receiver $r$ and the transmitter $t$, respectively, for the $s$-th ray, with $s=0,\ldots,S$. %
The index $s=0$ indicates the \ac{LOS} component of the channel. The angles $\theta$ and $\phi$ correspond to the horizontal and vertical directions, respectively. %
The term $A_s$ is the channel coefficient that contains the pathloss, shadowing and fast-fading. %
The channels between the receiver $r$ and transmitter $t$, $\mathbf{H}_{r,t}$, are defined as $\mathbf{H}_{\text{IRS},\text{BS}}$ = $\mathbf{J}$, $\mathbf{H}_{\text{\ac{UE}2},\text{BS}}$ = $\mathbf{H}_2$, $\mathbf{H}_{\text{\ac{UE}1},\text{IRS}}$ = $\mathbf{G}_1$ and $\mathbf{H}_{\text{\ac{UE}2},\text{IRS}}$ = $\mathbf{G}_2$. %

For the modeling of the steering vectors, the \ac{IRS} is designed as a \ac{URA}. The \acp{UE} and the \ac{BS} are considered to be equipped with horizontal \acp{ULA}, for which the angle $\phi$ is disregarded.  %

Due to the use of \ac{mmWave} bands, the pathloss, shadowing and Rician K-factor for the considered system are modeled according to \cite{3gpp.38.901} considering the \ac{UMa} scenario: 
\begin{equation}
  \text{PL} = 28 + 22\log_{10}\left(d_{3D}\right) + 20\log_{10}\left(f_c\right),
\end{equation}
in which $d_{3D}$ is the absolute distance between the transmitter and the receiver and $f_c$ is the carrier frequency. %
The shadow fading is modeled according to a log-normal distribution with standard deviation $\sigma=4$~dB. %
The Rician K-factor also follows a log-normal distribution, with $K\sim\mathcal{N}\left(9,3.5\right)$~dB~\cite{3gpp.38.901}, for all the links in the fig~\ref{FIG:single-cell}. %

The scattering in a \ac{UMa} scenario is considered to be rich, i.e., the channel has a large number of multi-paths. %
Therefore, all channels in the studied scenario are considered to have a full rank. %

\subsection{\acs{IRS} Phase Shift Setting}
Seeing that the main use of the \ac{IRS} in this work is to provide an alternative path for users under blockage, for the design of the~\ac{IRS} phase-shift, a~\ac{SVD} is performed on the channels $\mathbf{J}$ and $\mathbf{G}_1$. The phase-shift vector $\boldsymbol{\omega}$ is generated through the Hadamard product of the left singular vector of $\mathbf{J}$ associated with its highest singular value, $\mathbf{u}_{\mathbf{J}}$, and the right singular vector of $\mathbf{G}_1$ associated with its highest singular value,  $\mathbf{v}_{\mathbf{G}_1}^*$:
\begin{equation}
	\boldsymbol{\omega} = - \angle (\mathbf{u}_{\mathbf{J}} \odot \mathbf{v}_{\mathbf{G}_1}^*) \in \mathbb{C}^{N \times 1}\, ,
\end{equation} %
where $\odot$ denotes Hadamard product, and $*$ is the conjugate of a vector. %
\section{Proposed Precoding Design}\label{SEC:ProposedSolution}

Our proposed \ac{IRS}-aided network operates in two stages: i) There is the \ac{IRS} phase-shift design, also known as \ac{IRS} passive beamforming, and then ii) the digital precoder and combiner are computed. %

To support multi-user communication, we propose two \ac{BD}-based precoding schemes implemented at the \ac{BS}. On the \acp{UE} side, a traditional \ac{ZF} combiner matched to the intended channel is employed. %
In this study, all channel responses are assumed known at the~\ac{BS} by relying on the fact that practice they can be estimated, as, e.g., demonstrated in~\cite{Asim2023,Gil2023,Gomes2023}.

\subsection{General~\acs{BD} Framework}\label{SEC:Active_BF}
When it comes to \ac{MU-MIMO} interference mitigation techniques, \ac{BD} is well-known for its efficiency in maximizing either the throughput or the fairness of the system~\cite{Spencer2004}. %
\ac{BD}-based precoders have the potential to cancel interference toward non-intended users by introducing the following constraint:
\begin{equation}\label{EQ:BDstep1}
	\tilde{\mathbf{H}}_u\mathbf{F}_k = \mathbf{0}\strut_{N_{M,u}\times N_{M,u}} \,, \quad \forall u \neq k,
\end{equation}
where $u,k = 1,\dots,L$, represent \ac{UE} indexes, $L$ is the total number of \acp{UE} in the system, and $\tilde{\mathbf{H}}_l$ is the complementary channel of the $l$-th \ac{UE}, defined as:
\begin{equation}
	\tilde{\mathbf{H}}_l = \begin{bmatrix}\mathbf{H}^{\text{T}}_1 & \ldots & \mathbf{H}^{\text{T}}_{l-1} & \mathbf{H}^{\text{T}}_{l+1} & \ldots & \mathbf{H}^{\text{T}}_L \end{bmatrix}^{\text{T}}.
\end{equation}

In order to achieve the result in \eqref{EQ:BDstep1} and mix the signal for the intended users coherently, it is necessary for the precoder $\mathbf{F}_k$ to lie in the null-space of $\tilde{\mathbf{H}}_l$ and on the signal-space of $\mathbf{H}_k$. %
Both of which can be obtained by the~\ac{SVD} of $\tilde{\mathbf{H}}_l$ and $\mathbf{H}_k$, respectively. %
This technique  must obey the restriction that the number of transmitting antennas should be greater than or equal to the total number of receiving antennas. %

The precoder $\mathbf{F}_k$ is constructed as follows: 
\begin{equation}\label{EQ:BD_precoder_design}
  \begin{split}
    &	\tilde{\mathbf{H}}_l = \tilde{\mathbf{U}}_l\tilde{\mathbf{\Lambda}}_l \begin{bmatrix}\tilde{\mathbf{V}}^{\strut\text{non-zero}}_l & \tilde{\mathbf{V}}^{\strut\text{zero}}_l\end{bmatrix}^\text{H},  \\
              & \mathbf{H}_k  \tilde{\mathbf{V}}^{\strut\text{zero}}_l= \mathbf{U}_k\mathbf{\Lambda}_k \begin{bmatrix}\mathbf{V}^{\strut\text{non-zero}}_k & \mathbf{V}^{\strut\text{zero}}_k\end{bmatrix}^\text{H},  \\
    & \mathbf{F}_k= \tilde{\mathbf{V}}^{\strut\text{zero}}_l\mathbf{V}^{\strut\text{non-zero}}_k.
  \end{split}
\end{equation}

As shown in~\cite{Ning2022}, the classic \ac{BD} technique is not feasible for \ac{IRS}-aided scenarios, given that the \acp{UE} channels and the \ac{IRS} phase-shift vectors are coupled, 
which leads to the right singular vectors of $\tilde{\mathbf{H}}_l$ and $\mathbf{H}_k$ not being disjoint. %
In this context, two techniques are proposed in the sequel to overcome this issue. %
\subsection{\ac{PIB} precoder design}\label{SEC:PIRSBD}

In this first approach, the \ac{IRS} precise beamforming~\cite{zappone2021} is taken into account and it is assumed that the beam leakage is minimal. %
Therefore, for the system presented in~\FigRef{FIG:single-cell}, the \ac{IRS} leakage channel $\mathbf{G}_{2}\boldsymbol{\Omega}\mathbf{J}$ is neglected in the precoder design and $\bar{\mathbf{H}}_1 = \mathbf{G}_{1}\boldsymbol{\Omega}\mathbf{J}$ and $\mathbf{H}_{2}$ are the only contemplated channels, %
which can be classically block diagonalized by considering the complementary channels as follows: %
\begin{equation}
  \begin{split}\label{EQ:PIB_Complementary}
    & \tilde{\mathbf{H}}_1 = \mathbf{H}_{2}, \\
    & \tilde{\mathbf{H}}_2 = \mathbf{G}_{1}\boldsymbol{\Omega}\mathbf{J}.
  \end{split}
\end{equation}
Given the complementary channels in~\eqref{EQ:PIB_Complementary}, the precoder can be design using~\eqref{EQ:BD_precoder_design}.

The assumption that $\mathbf{G}_{2}\boldsymbol{\Omega}\mathbf{J}$ is negligible compared to $\mathbf{H}_2$ makes this method suitable for interference mitigation in the considered scenario. The veracity of this assumption will be further analyzed in~\SecRef{SEC:SimRes}. %
\subsection{\ac{FIB} precoder design}

Unlike the previous method, this second technique considers in its design the contributions of  $\mathbf{G}_{2}\boldsymbol{\Omega}\mathbf{J}$ on both useful signal and interference components of~\eqref{EQ:y2CompleteEquation}.
To do that, instead of considering the channel $\bar{\mathbf{H}}_2 = \mathbf{H}_2 + \mathbf{G}_2\boldsymbol{\Omega}\mathbf{J}$ of~\eqref{EQ:MIMOMatricial_step}, these components are treated as two separate independent channels of \ac{UE}2.
Based on this consideration, the complementary channels become
\begin{equation}\label{EQ: FIRSBD_complementary_channels}
  \begin{split}
    & \tilde{\mathbf{H}}_1 = \begin{bmatrix}\mathbf{H}^{\text{T}}_{2}  & \left(\mathbf{G}_{2}\boldsymbol{\Omega}\mathbf{J}\right)^{\text{T}}  \end{bmatrix}^{\text{T}} \, , \\
    & \tilde{\mathbf{H}}_2 = \mathbf{G}_{1}\boldsymbol{\Omega}\mathbf{J}.
  \end{split}
\end{equation}

It is important to notice that, because the \ac{IRS} channels are coupled, i.e., they depend on the~\ac{IRS} phase-shift vector, the \ac{BD} technique is not able to fully cancel all the interfering signals~\cite{Ning2022}. %
Thus, in the configuration of~\eqref{EQ: FIRSBD_complementary_channels}, the signals traversing these coupled channels can be completely canceled after applying the precoder designed with~\eqref{EQ:BD_precoder_design} at the cost of receiving some residual interference from the direct channel $\mathbf{H}_{2}$. %

\subsection{Combiner design}

For both precoding techniques, it is considered that the \acp{UE} are employing traditional \ac{ZF} combining on the strongest channel, which can be expressed as:
\begin{equation}
  \begin{split}
    & \mathbf{W}_1 = \begin{bmatrix}\mathbf{G}_{1}\boldsymbol{\Omega}\mathbf{J}\mathbf{F}_{1}\end{bmatrix}^\dagger ,\\
    & \mathbf{W}_2 = \begin{bmatrix}\mathbf{H}_{2}\mathbf{F}_{2}\end{bmatrix}^\dagger,
  \end{split}
\end{equation}
where $(\cdot)^\dagger$ represents the Moore-Penrose pseudo-inverse.

\section{Simulations Results}\label{SEC:SimRes}

In this section, the performance of each of the proposed methods will be evaluated in terms of~\ac{SE} and compared with i) the solution presented in~\cite{Ning2022}, and with ii) a block-diagonalized~\ac{MU-MIMO} system without~\ac{IRS} that considers a (hypothetical) strong direct path between~\ac{BS} and~\ac{UE}1, then acting as a benchmark.

The scenario illustrated in~\FigRef{FIG:single-cell} was simulated considering $M=32$ antennas at the~\ac{BS}, $N=64$ reflecting elements at the~\ac{IRS}, $P=Q=8$ antennas at the~\acp{UE}, and the number of \ac{RF} chains in both \acp{UE} was set to 2, which leads to $N_s=N_s^{(1)}=N_s^{(2)}=2$ streams. The other parameters considered to simulate the studied scenario are exposed in~\TabRef{table:sim_parameters}, where $d_{2D}$ represents the horizontal distance. %
\begin{table}[ht]
  \centering
  \caption{Simulation parameters.}\label{table:sim_parameters} 
  \begin{tabularx}{\columnwidth}{XX}
    \toprule
    \textbf{Parameter} & \textbf{Value} \\
    \midrule
    Carrier Frequency &  $f_c$ = 28 GHz \\
    Antenna Spacing & $d$ = $\lambda/2$ m \\
    ${\text{UE height}}$ & 1.5m \\
    ${\text{IRS height}}$ & 8m \\
    ${\text{BS height}}$ & 25m \\
    $d_{2D}$ between ~\ac{BS} and ~\ac{IRS} & 100m \\
    $d_{2D}$ between ~\ac{BS} and ~\acp{UE} & 100m \\
    $d_{2D}$ between ~\ac{IRS} and ~\ac{UE}1 & 51.7m \\
    $d_{2D}$ between ~\ac{IRS} and ~\ac{UE}2 & 100m \\
    Number of channel realizations & 10000 \\
    Noise power & $\sigma_1^2=\sigma_2^2=-80$ dBm \\
    \bottomrule
  \end{tabularx}  
\end{table}

\begin{figure}[htb]
  \centering
  \includegraphics[width=0.9\columnwidth]{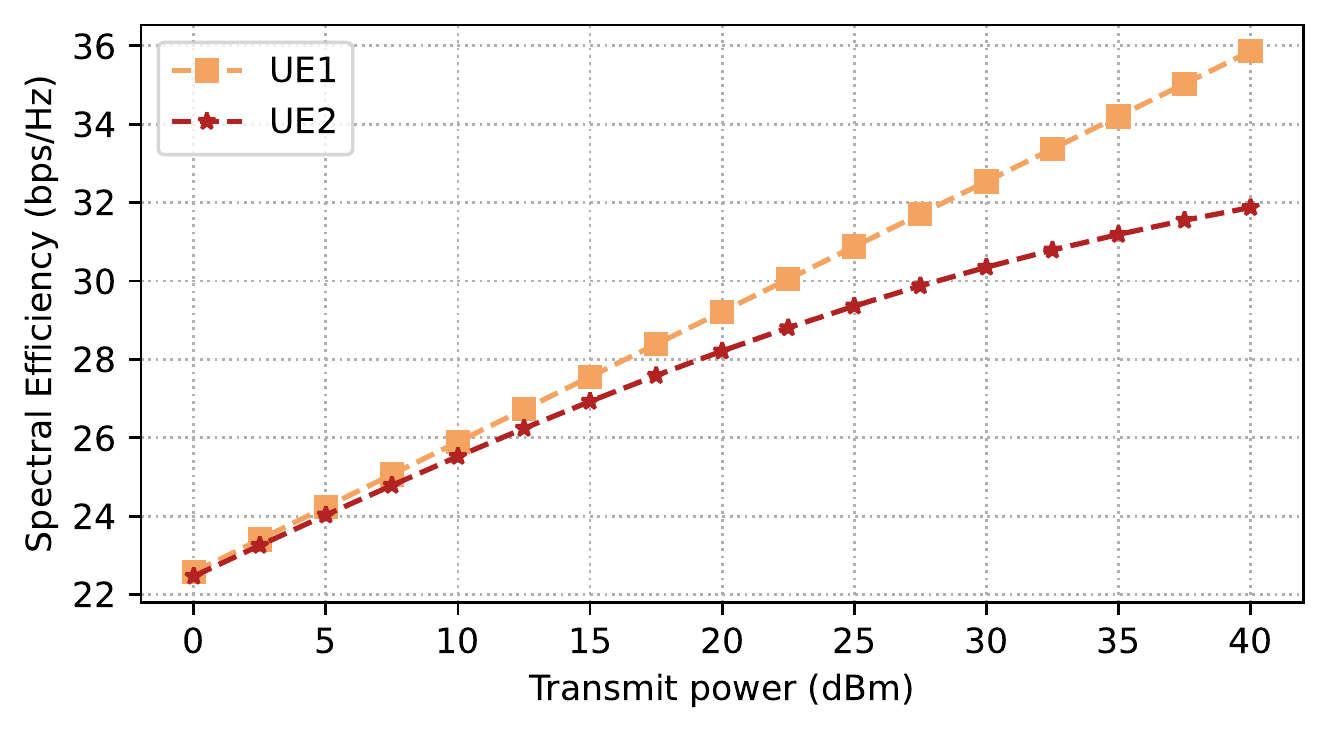}
  \caption{SE vs \ac{BS} total transmit power for~\ac{PIB} precoder design.}
  \label{FIG:PIRSBD_wi}
\end{figure}
\FigRef{FIG:PIRSBD_wi} presents the~\ac{SE} for each~\ac{UE} as a function of the transmit power employed at the~\ac{BS} when the~\ac{PIB} method is applied in the system.%
It can be noticed that the~\ac{SE} for~\ac{UE}1 grows linearly, which is expected given that both~\acp{UE} are in the high~\ac{SNR} regime and because this technique is able to fully cancel signals flowing through $\mathbf{G}_{1}\boldsymbol{\Omega}\mathbf{J}\mathbf{F}_{2}$. %
In contrast to it, the~\ac{UE}2 presents two behaviors: until 15~\si{dBm}, the~\ac{SE} of \ac{UE}2 grows linearly, similarly to~\ac{UE}1, and after that it presents a still increasing behavior but with a lower slope. %
This  is also expected since this technique does not cancel the interference arriving from channel $\mathbf{G}_{2}\boldsymbol{\Omega}\mathbf{J}$ and as the power of the~\ac{BS} increases so does the interfering signal coming through $\mathbf{G}_{2}\boldsymbol{\Omega}\mathbf{J}$. %

Considering the assumptions made in~\SecRef{SEC:PIRSBD} and based on this result, we can conclude that if the interfering channel has low gain, the beam leakage can be considered minimal and be neglected in the precoder design. %

\begin{figure}[bt]
	\centering
	\includegraphics[width=0.9\columnwidth]{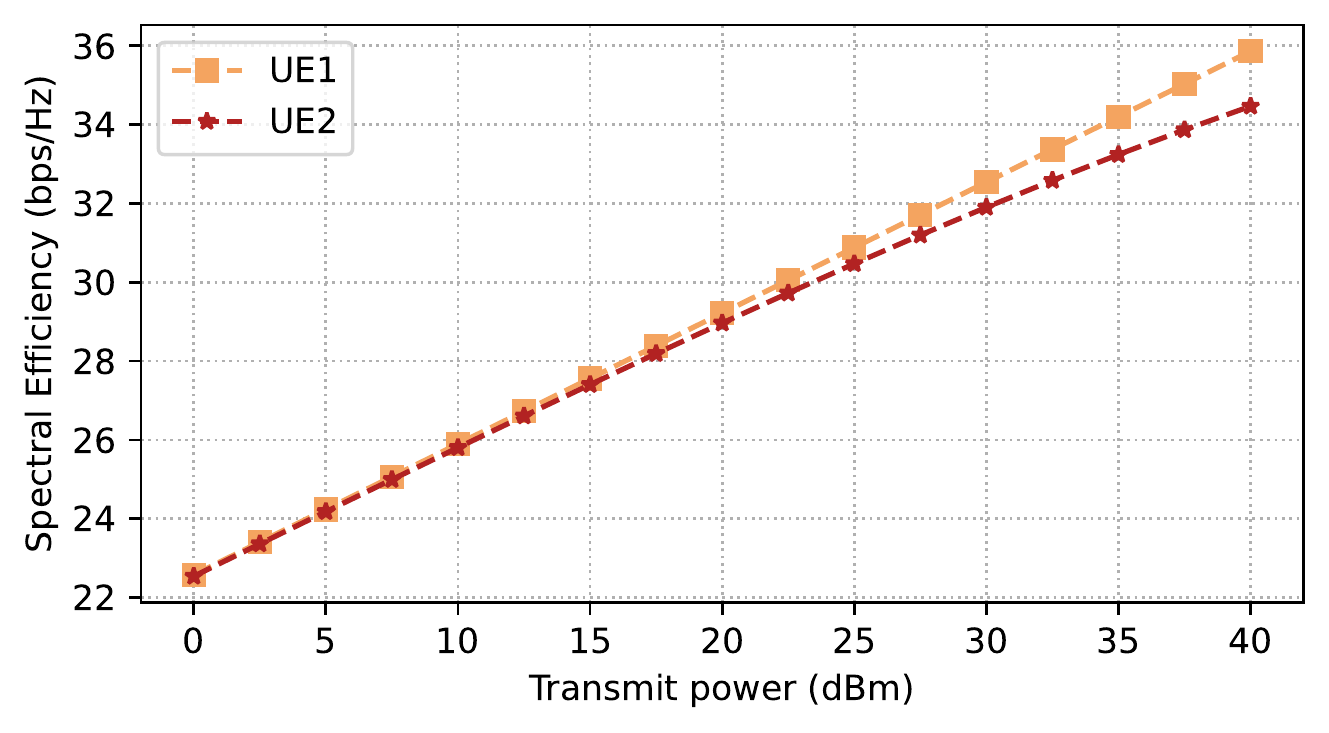}
	\caption{SE vs \ac{BS} total transmit power for~\ac{FIB} precoder design.}
	\label{FIG:FIRSBD_wi}
\end{figure}
\FigRef{FIG:FIRSBD_wi} exhibits the~\ac{SE} for each~\ac{UE} as a function of the transmit power employed at the~\ac{BS} when the~\ac{FIB} method is applied on the system. %
As expected, the~\ac{FIB} technique outperforms the~\ac{PIB}, since it is more capable of canceling interference arriving through $\mathbf{G}_{2}\boldsymbol{\Omega}\mathbf{J}$ and even thought the~\ac{UE}2 has some residual interference arriving through $\mathbf{H}_2$, it does not impact significantly on its performance. %

In the following, a comparison between the no-\ac{IRS}~\ac{MU-MIMO} and the~\ac{IRS}-aided~\ac{MU-MIMO} will be provided. %
In the no-\ac{IRS}~\ac{MU-MIMO}, all system parameters are kept the same and the only modification occurs on~\ac{UE}1, whose~\ac{IRS} is replaced by a direct channel with strong~\ac{LOS} similarly to~\ac{UE}2. %
In this scenario the~\ac{BS} precoder is designed following the classical~\ac{BD} method. 

\begin{figure}[t]
	\centering
	\includegraphics[width=0.9\columnwidth]{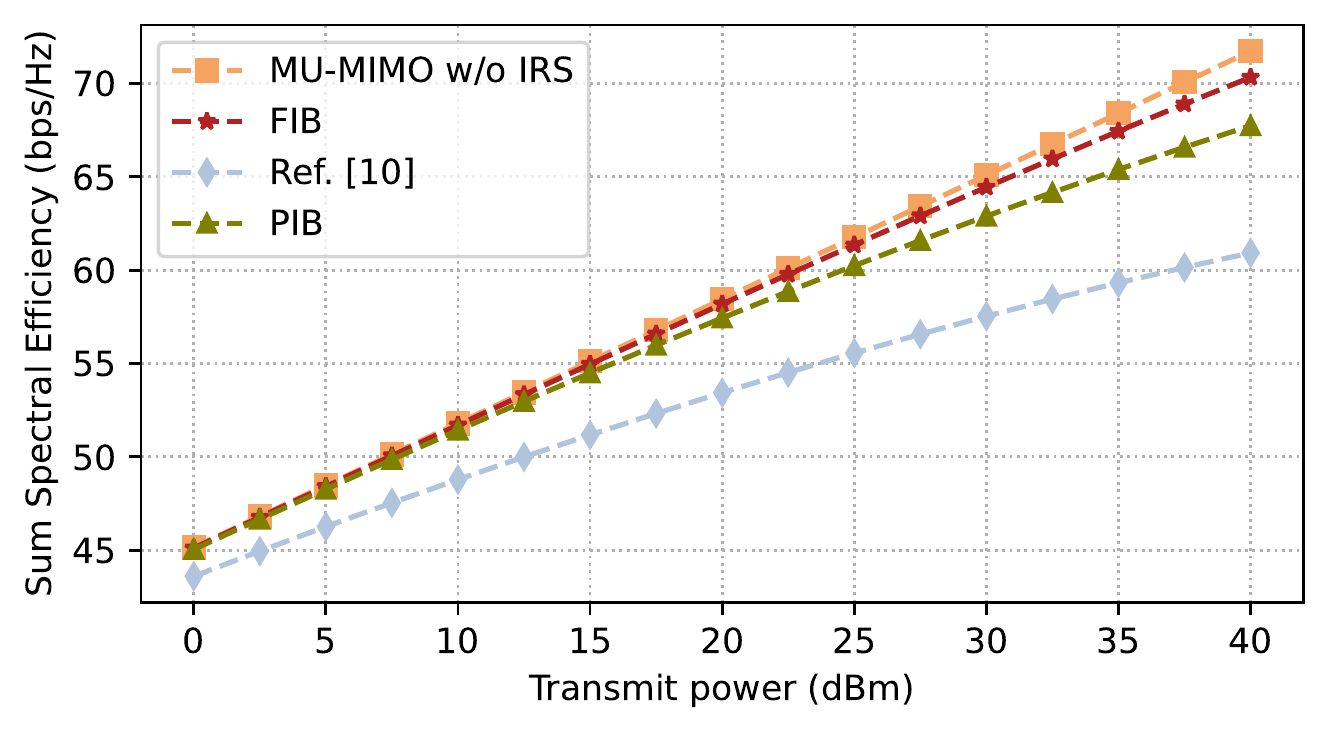}
	\caption{SE vs \ac{BS} transmit power for all compered precoder design.}
	\label{FIG:SumSE}
\end{figure}

\FigRef{FIG:SumSE} shows the sum~\ac{SE} as a function of the transmit power at the~\ac{BS} for each method presented, as well as for an adapted version of the precoder from~\cite{Ning2022}. %
Therein, only one user performs interference cancellation, since its proposal needs one~\ac{IRS} per user. %
An additional curve for the~\ac{MU-MIMO} without~\ac{IRS} is also provided, in which the~\ac{IRS} link is replaced by a direct link with the same propagation properties of $\mathbf{H}_2$ for comparison purposes. %

It can be observed that the no-\ac{IRS} setup presents the highest~\ac{SE}. This is due the considered direct link for~\ac{UE}1.
However, in the cases where that link is not available, as defined in Section~\ref{SEC:SystemModel}, the \ac{IRS}~\ac{MU-MIMO} with the~\ac{FIB} method presents the best results, provided that it is the most robust method. %
However, the overall performance of the three scenarios does not significantly differs from each other. %
In contrast, the method in~\cite{Ning2022} has the worst performance, since its solution need one~\ac{IRS} per user. %
Hence, one user can fully cancel the interference and the other has unmitigated interference arriving from the~\ac{IRS}.

As for computational complexity, the precoder design method of~\cite{Ning2022} presents the highest complexity, since it cancels the interference on \ac{BS}-\ac{IRS} link instead of \ac{BS}-\ac{IRS}-\ac{UE} link, requiring an~\ac{SVD} of higher dimension for the precoder design. %
It is followed by the~\ac{FIB} method, which uses the direct link and the \ac{BS}-\ac{IRS}-\ac{UE} link. %
The~\ac{PIB} is the least complex method herein, since it uses only the \ac{BS}-\ac{IRS}-\ac{UE} link in its formulations.

It is worth mentioning that, in an \ac{IRS}-aided network, as the~\ac{PIB} method performs similarly to the~\ac{FIB} one and outperforms the method of~\cite{Ning2022}, but with reduced complexity.
The~\ac{PIB}  may be useful in scenarios in which high spatial multiplexing gains are achievable, directional antennas are employed at the receiver, the interfering channel suffers from blockage, and in many other cases where the interfering channel is negligible compared to the direct one.

\section{Conclusion}\label{SEC:Conclusion}

In this paper, the impact of multi-user interference in~\ac{IRS}-aided networks was studied. %
Two precoding methods based on~\ac{BD} were proposed to spatially orthogonalize users' signals. %
It was observed that both methods perform well when the interfering channels created by the IRS has low gain, especially~\ac{PIB}, which has the same limitations of the original~\ac{BD} technique. %
The~\ac{FIB} technique, even though being more demanding in computational and antenna resources, grants better interference cancellation and performs closely  to the no-\ac{IRS}~\ac{MU-MIMO} case, since the former not only better orthogonalizes the users but also takes advantage of the additional path provided by the~\ac{IRS} for the non-intended users. %
Future works include the study of multi-cell and multi-IRS scenarios, other mitigation interference methods, joint~\ac{IRS} phase shift and precoder optimization and~\ac{IRS} splitting for serving multiple users.

\printbibliography

\end{document}